\documentclass[conference,a4paper]{IEEEtran}
\IEEEoverridecommandlockouts

\usepackage[hidelinks]{hyperref}
\usepackage{orcidlink}

\usepackage{dblfloatfix}
\usepackage{graphicx}
\graphicspath{{Figures/}}
\usepackage{placeins}

\usepackage{booktabs}
\usepackage{makecell}
\usepackage{multirow}
\usepackage[usestackEOL]{stackengine}
\usepackage{tcolorbox}
\usepackage[table]{xcolor}

\usepackage{minted}
\setminted{
  fontsize=\scriptsize,
  style=friendly,
  bgcolor=gray!10,
  baselinestretch=0.9,
  breaklines,
  breakanywhere,
  breakautoindent,
  breakindent=1em,
  xleftmargin=0.3em,
  tabsize=2,
  autogobble
}

\usepackage[numbers,compress]{natbib}

\usepackage[edges]{forest}
\forestset{
  default preamble={
    for tree={
      grow'=0,
      draw,
      thick,
      align=center,
      edge={thick},
      forked edge,
      l sep=0.4cm,
      s sep=0.3cm,
      minimum height=2.2em,
      inner sep=3pt,
      anchor=west,
      font=\sffamily\scriptsize,
    }
  }
}

\begin{document}

\title{\uppercase{Earth Embeddings as Products: Taxonomy, Ecosystem, and Standardized Access}}

\author{
    \IEEEauthorblockN{
    Heng Fang\IEEEauthorrefmark{2} \IEEEauthorrefmark{1} \orcidlink{0009-0009-9256-7306} \qquad
    Adam J. Stewart\IEEEauthorrefmark{3} \IEEEauthorrefmark{1} \orcidlink{0000-0002-0468-5006} \qquad
    Isaac Corley\IEEEauthorrefmark{4} \orcidlink{0000-0002-9273-7303} \\
    Xiao Xiang Zhu\IEEEauthorrefmark{3} \orcidlink{0000-0001-5530-3613} \qquad
    Hossein Azizpour\IEEEauthorrefmark{2} \orcidlink{0000-0001-5211-6388}
    }

    \vspace{0.5em}

    \IEEEauthorblockA{
        \IEEEauthorrefmark{2} \textit{KTH Royal Institute of Technology, Stockholm, Sweden} \\
        \IEEEauthorrefmark{3} \textit{Chair of Data Science in Earth Observation, Technical University of Munich, Munich, Germany} \\
        \IEEEauthorrefmark{4} \textit{Wherobots, San Antonio, USA}
    }

    \vspace{0.5em}
    
    \IEEEauthorblockA{\{hfang, azizpour\}@kth.se \qquad \{adam.stewart, xiaoxiang.zhu\}@tum.de \qquad isaac@wherobots.com}

    \thanks{\IEEEauthorrefmark{1} These authors contributed equally to this work.}
}


\maketitle

\begin{abstract}
Geospatial Foundation Models (GFMs) provide powerful representations, but high compute costs hinder their widespread use. Pre-computed embedding data products offer a practical ``frozen'' alternative, yet they currently exist in a fragmented ecosystem of incompatible formats and resolutions. This lack of standardization creates an engineering bottleneck that prevents meaningful model comparison and reproducibility. We formalize this landscape through a three-layer taxonomy: Data, Tools, and Value. 
We survey existing products to identify interoperability barriers. To bridge this gap, we extend TorchGeo with a unified API that standardizes the loading and querying of diverse embedding products. By treating embeddings as first-class geospatial datasets, we decouple downstream analysis from model-specific engineering, providing a roadmap for more transparent and accessible Earth observation workflows.
\end{abstract}


\section{Introduction}
\label{sec:introduction}

Recent years have witnessed a surge in Geospatial Foundation Models (GFMs), task-agnostic models pre-trained on vast data sources capable of producing low-dimensional fixed-length vectors that capture semantic, spatial, and temporal information~\cite{zhu2024foundations}. While models like AlphaEarth~\cite{alphaearth} and Clay~\cite{clay} promise powerful representations, the ecosystem for their derived products, \textit{Earth Embeddings}, remains unstructured. The field is rapidly emerging, with approximately 20---25 distinct works appearing in recent literature. While recent position papers have provided valuable perspectives on the potential and research roadmap of these representations~\cite{gomes2025lossy, rolf2025earth, klemmer2025earth}, they primarily address the theoretical landscape. A systematic review focused on the operational ecosystem is still absent.

Consequently, for practitioners, the barrier to entry remains high. The current landscape is fragmented, with embedding products differing significantly in file formats, spatial resolutions, and licensing terms. There is no unified methodology to load or evaluate them.

To address this fragmentation, we present a unified framework that bridges the gap between theoretical vision and practical access. We identify the lack of a common interface as the primary bottleneck and resolve this through integration with the TorchGeo library~\cite{torchgeo}. Our specific contributions are:
\begin{itemize}
    \item \textbf{A Comprehensive Survey}: We organize seven existing embedding products into a structured taxonomy and provide a detailed metadata atlas (resolution, license, etc.).
    \item \textbf{Unified Integration}: We implement standardized data loaders for these embeddings in TorchGeo, enabling users to access diverse embeddings via a single API.
\end{itemize}

\begin{figure}[t]
    \centering
    \resizebox{0.9\columnwidth}{!}{
      \begin{forest}
        [The Earth\\Embedding\\Landscape, minimum width=2.2cm, 
          [Embeddings\\(Data), minimum width=2.2cm
            [Location\\Embeddings, minimum width=2.8cm]
            [Patch\\Embeddings, fill=gray!10, minimum width=2.8cm]
            [Pixel\\Embeddings, fill=gray!10, minimum width=2.8cm]
          ]
          [Analysis\\Frameworks\\(Tools), minimum width=2.2cm
            [Benchmarks, minimum width=2.8cm]
            [Open\\Challenges, minimum width=2.8cm]
          ]
          [Downstream\\Applications\\(Value), minimum width=2.2cm
            [Mapping, minimum width=2.8cm]
            [Retrieval, minimum width=2.8cm]
          ]
        ]
      \end{forest}
    }
    \caption{\textbf{Taxonomy of the Earth Embedding Landscape.} We organize the ecosystem into three functional layers: Data, Tools, and Value. Highlighted blocks indicate the specific scope of the assessment in Section~\ref{sec:ecosystem}.}
    \label{fig:taxonomy}
\end{figure}
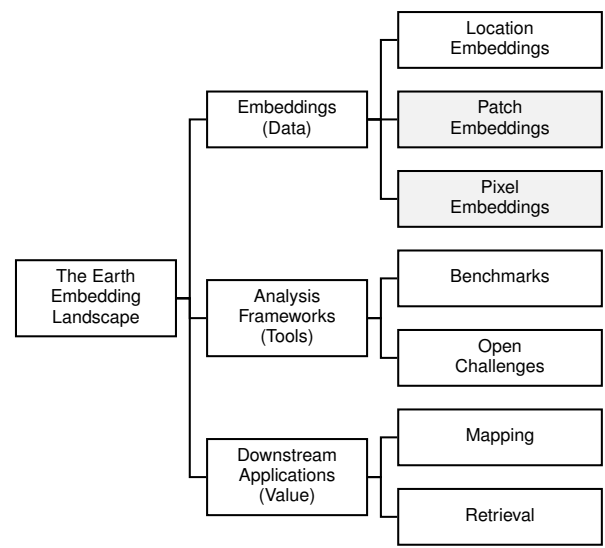

\section{The Earth Embedding Landscape}\label{sec:landscape}

\begin{table*}[htbp]
    \centering
    \caption{\textbf{Embedding products available as of December 2025.} Global coverage only implies land surfaces, no existing products cover oceans. Temporal resolution is divided into ``snapshot'' for embeddings generated from a single mosaic and ``annual'' for embeddings generated from annual time series data. *Product has sparse spatial or temporal coverage.}
    \begin{tabular}{lclrlcrcr}
        \toprule
        & & \multicolumn{2}{c}{\textbf{Spatial}} & \multicolumn{2}{c}{\textbf{Temporal}} \\
        \cmidrule(l){3-4} \cmidrule(l){5-6}
        \textbf{Product} & \textbf{Kind} & \textbf{Extent} & \textbf{Resolution} & \makecell[c]{\textbf{Extent}} & \textbf{Resolution} & \textbf{Dimensions} & \textbf{Dtype} & \textbf{License} \\
        \midrule
        Clay Embeddings~\cite{clay} & Patch & Global* & 5.12~km & 2018--2023* & Snapshot & 768 & float32 & ODC-By-1.0 \\
        Major TOM Embeddings~\cite{majortom} & Patch & Global & 2.14--3.56~km & 2015--2024* & Snapshot & 2048 & float32 & CC-BY-SA-4.0 \\
        Earth Index Embeddings~\cite{earthindex} & Patch & Global & 320~m\phantom{k} & \phantom{2024--}2024 & Snapshot & 384 & float32 & CC-BY-4.0 \\
        Copernicus-Embed~\cite{copernicus} & Patch & Global & 0.25\textdegree\phantom{km} & \phantom{2021--}2021 & Annual & 768 & float32 & CC-BY-4.0 \\
        \midrule
        Presto Embeddings~\cite{prestoembed} & Pixel & Togo & 10~m\phantom{k} & 2019--2020 & Annual & 128 & uint16 & CC-BY-4.0 \\
        Tessera Embeddings~\cite{tessera} & Pixel & Global* & 10~m\phantom{k} & 2017--2025* & Annual & 128 & int8 \(\rightarrow\) float32 & CC0-1.0 \\
        Google Satellite Embedding~\cite{alphaearth} & Pixel & Global & 10~m\phantom{k} & 2017--2025 & Annual & 64 & int8 \(\rightarrow\) float64 & CC-BY-4.0 \\
        \bottomrule
    \end{tabular}
    \label{tab:embeddings}
\end{table*}

\subsection{Distinction: Foundation Models vs. Embedding Products}

To clarify our scope, we first distinguish between the \textit{generator} (the model) and the \textit{asset} (the generated embedding).

\textbf{Foundation Models (Dynamic Inference)}: These architectures are distributed as public weights (e.g., DOFA~\cite{dofa}, OlmoEarth~\cite{olmoearth}). While they serve as embedding generators, utilizing them requires an active inference pipeline. Users must manage data preprocessing and leverage GPU resources to compute representations from raw imagery. This offers flexibility but imposes hardware and engineering barriers.

\textbf{Embedding Products (Static Data)}: In contrast, embedding products (e.g., Google Satellite Embedding~\cite{alphaearth}) are distributed as static, pre-computed vector archives. These are ``frozen'' assets. They decouple the representation from the computational cost of generation, allowing users to conduct analysis directly without accessing the original model.

\begin{tcolorbox}[
  colback=gray!10,
  colframe=black!40,
  boxrule=0.4pt,
  arc=2pt,
  left=5pt,
  right=5pt,
  top=5pt,
  bottom=5pt
]
\footnotesize
\textbf{Why this matters}: In practice, this distinction is not merely academic. Treating a static embedding product as a proxy for a foundation model allows for efficient application but limits generalizability. Confusing the two often leads to invalid performance claims, as products are constrained by the specific data snapshot they were computed on.
\end{tcolorbox}

\subsection{Taxonomy of Earth Embeddings}
Based on our curated community repository \href{https://github.com/hfangcat/Awesome-Geospatial-Embeddings}{\texttt{Awesome-Geospatial-Embeddings}}, we structure the current ecosystem into three functional layers, as visualized in Fig.~\ref{fig:taxonomy}:




\subsubsection{Embeddings (Data)}

This layer encompasses the pre-computed representations themselves. To clarify their structural properties and intended utility, we categorize them by their \textit{spatial granularity}:

\textbf{Location Embeddings}: Representations that encode spatiotemporal coordinates \(\{x, y, t\}\) independent of imagery. Models like SatCLIP~\cite{klemmer2025satclip} and LocationEncoder~\cite{russwurm2023geographic} map global latitude/longitude coordinates to latent vectors, capturing environmental contexts (e.g., climate zones).

\textbf{Patch-level Embeddings}: Single vector representations summarizing an entire image chip (e.g., \(256 \times 256\) pixels). Products in this category, such as Clay~\cite{clay} or Major TOM~\cite{majortom} embeddings, compress high-dimensional sensor data into compact global descriptors. These are optimized for identifying semantic similarity, making them the fundamental unit for \textit{search and retrieval} tasks.

\textbf{Pixel-level Embeddings}: Spatially preserved feature maps where each pixel is assigned a vector, with or without context of neighboring pixels. Unlike global descriptors, these embeddings retain local structural information. They are essential for fine-grained \textit{mapping} tasks, such as semantic segmentation, where spatial precision is paramount.

\subsubsection{Analysis Frameworks (Tools)} 

To understand the utility and behavior of these representations, the community has established specific analytical resources:

\textbf{Benchmarks \& Evaluation:} Standardized frameworks are emerging to probe embedding quality. Examples include NeuCo-Bench~\cite{vinge2025neuco} for general vision-based embedding evaluation and GeoINRID~\cite{rao2025measuring} for measuring the intrinsic dimension of coordinate-based embeddings.

\textbf{Open Challenges:} Competitions such as the Embed2Scale Challenge (CVPR 2025)~\cite{albrecht20252025} and TerraMind Blue-Sky Challenge~\cite{jakubik2025terramind} play a crucial role in driving standardization and testing embeddings under compression or retrieval constraints.

\subsubsection{Downstream Applications (Value)}

Finally, the landscape demonstrates the practical utility of frozen embeddings. While most research focuses on \textbf{mapping} (e.g., cropland mapping~\cite{prestoembed}, poverty mapping~\cite{pettersson2025leveraging}), we find a clear split in utility: dense \textit{pixel-level} features excel at fine-grained segmentation, whereas sparse \textit{patch-level} vectors are ideal for \textbf{retrieval}, an emerging area with significant potential.

\section{The Embedding Product Ecosystem}\label{sec:ecosystem}


Excluding foundation models and location encoders that require the user to run the model themselves to generate new embeddings, we explore all known Earth embedding products listed in Table~\ref{tab:embeddings} and compare them on their spatial and temporal extent and resolution, compression, and license. We also provide an overview of the work that went into producing these embeddings, including the model architecture, pre-training technique, and data, and comment on the reproducibility of these embeddings in Table~\ref{tab:reproducibility}.

\begin{table*}[htbp]
    \centering
    \caption{\textbf{Reproducibility of embedding products available as of December 2025.} Models are broken down into the model architecture and the self-supervised learning method used to pre-train the model weights. Data is broken down into the data used during pre-training and the data used during inference to generate the embeddings. Licenses are broken down into the license for the model code, model weights, and inference data. NASA/USGS data tends to be public domain, ESA data tends to be under the Copernicus Terms \& Conditions, and JAXA data has its own Terms of Service. Highlighted models are not open source.}
    \begin{tabular}{@{}cccccccc@{}}
        \toprule
        & \multicolumn{2}{c}{\textbf{Model}} & \multicolumn{2}{c}{\textbf{Data}} & \multicolumn{3}{c}{\textbf{Licenses}} \\
        \cmidrule(l){2-3} \cmidrule(l){4-5} \cmidrule(l){6-8}
        \textbf{Product} & \textbf{Architecture} & \textbf{Training} & \textbf{Training} & \textbf{Inference} & \textbf{Code} & \textbf{Weights} & \textbf{Data} \\
        \midrule
        \multirow{3}{*}{\makecell{Clay \\ Embeddings}} & & & Landsat~8/9, NAIP, MODIS & & & & public domain \\
        & Clay & MAE & Sentinel~2 & Sentinel~2 & Apache-2.0 & Apache-2.0 & Copern.~T\&C \\
        & & & LINZ & & & & CC-BY-4.0 \\
        \midrule
        \multirow{7}{*}{\makecell{Major TOM \\ Embeddings}} & ResNet-50 & DINO & Sentinel~2 & Sentinel~2 & Apache-2.0 & CC-BY-4.0 & CC-BY-SA-4.0 \\
        & ResNet-50 & MoCo~v2 & Sentinel~1 & Sentinel~1 & Apache-2.0 & CC-BY-4.0 & CC-BY-SA-4.0 \\
        & DINOv2-L & DINOv2 & LVD-142M & Sentinel~2 (RGB) & Apache-2.0 & Apache-2.0 & CC-BY-SA-4.0 \\
        & ViT-SO400M & SigLIP & WebLI & Sentinel~2 (RGB) & Apache-2.0 & Apache-2.0 & CC-BY-SA-4.0 \\
        & ResNet-50 & DeCUR & Sentinel~1/2 & Sentinel~2 & Apache-2.0 & Apache-2.0 & CC-BY-SA-4.0 \\
        & ResNet-50 & DeCUR & Sentinel~1/2 & Sentinel~2 & Apache-2.0 & Apache-2.0 & CC-BY-SA-4.0 \\
        & ConvNeXt~v2 & MP-MAE & MMEarth & Sentinel~2 & MIT & \cellcolor{gray!10}{CC-BY-NC-4.0} & CC-BY-SA-4.0 \\
        \midrule
        \makecell{Earth Index \\ Embeddings} & DINOv2-S & SoftCon & Sentinel~2 & Sentinel~2 & Apache-2.0 & CC-BY-4.0 & CC-BY-4.0 \\
        \midrule
        \makecell{Copernicus-\\Embed} & \makecell{Copernicus-\\FM} & \makecell{MAE \\ Distillation} & \makecell{Sentinel~1/2/3/5P \\ Copernicus DEM} & \makecell{Sentinel~1/2/3/5P \\ Copernicus DEM} & Apache-2.0 & CC-BY-4.0 & CC-BY-4.0 \\
        \midrule
        \makecell{Presto \\ Embeddings} & Presto & MAE & \makecell{Sentinel~1/2, ERA5 \\ Dynamic World} & \makecell{Sentinel~1/2, ERA5 \\ SRTM} & MIT & MIT & \makecell{Copern.~T\&C \\ CC-BY-4.0} \\
        \midrule
        \makecell{Tessera \\ Embeddings} & Tessera & \makecell{Barlow \\ Twins} & Sentinel~1/2 & Sentinel~1/2 & MIT & CC0-1.0 & Copern.\ T\&C \\
        \midrule
        \multirow{5}{*}{\makecell{Google \\ Satellite \\ Embedding}} & \multirow{5}{*}{\makecell{AlphaEarth \\ Foundations}} & & Sentinel, Copern.~DEM, ERA5 & Sentinel~1/2 & \cellcolor{gray!10} & \cellcolor{gray!10} & Copern.~T\&C \\
        & & Contrastive & Landsat, GEDI, GRACE, NLCD & Landsat~8/9 & \cellcolor{gray!10} & \cellcolor{gray!10} & public domain \\
        & & MAE & ALOS PALSAR ScanSAR & & \cellcolor{gray!10}Proprietary & \cellcolor{gray!10}Proprietary & JAXA ToS \\
        & & Distillation & Wikipedia articles & & \cellcolor{gray!10} & \cellcolor{gray!10} & CC-BY-SA-4.0 \\
        & & & GBIF & & \cellcolor{gray!10} & \cellcolor{gray!10} & CC-BY-4.0 \\
        \bottomrule
    \end{tabular}
    \label{tab:reproducibility}
\end{table*}


In compiling this list, we identified three trends largely following trends within the broader computer vision and self-supervised learning (SSL) communities:

\textbf{Patch \(\rightarrow\) pixel}: Early embedding efforts focused on generating a single high-dimensional embedding for each image patch due to computational requirements. These patch embeddings are typically stored in vector file formats like GeoParquet and are well suited to spatial- and embedding-based search/retrieval tasks. More recent efforts have focused on pixel embeddings, with or without spatial context. Pixel embeddings are typically lower dimensional, and practitioners have experimented with various compression and quantization strategies for reducing storage requirements.

\textbf{Model architectures}: Early embeddings were primarily generated using off-the-shelf convolutional architectures like ResNet~\cite{resnet} and ConvNeXt~\cite{convnext}, while more recent embeddings are primarily generated by custom transformer architectures. These architectures make it easier to incorporate spatial, temporal, and spectral metadata, and better support multi-modal input embedding~\cite{dofa} at the cost of significant engineering time to operationalize and deploy.

\textbf{SSL techniques}: We see a trend from simpler contrastive learning techniques, to more modern masked autoencoding (MAE) and distillation techniques. The large batch sizes and data augmentation required by contrastive learning can be prohibitively expensive for training large models. In contrast, MAEs are simple and efficient, supporting larger models trained on larger datasets. Following the success of DINO, student--teacher distillation is now making a comeback.

While Earth embeddings have come a long way, our analysis also revealed three major adoption barriers limiting ease-of-use and cross-comparison of different embedding products:

\textbf{Distribution:} The biggest challenge to cross-comparison is inconsistency in embedding distribution. Embeddings are distributed across Source Cooperative (Clay, Earth Index), Hugging Face (Major TOM, Copernicus-Embed), Google Earth Engine (Google, Presto), and private servers (Tessera). Patch-based embeddings, while often created from raster grids, are typically distributed in vector file formats like GeoParquet. These make streaming easy for search/retrieval applications, but are more challenging to rasterize for low-resolution land cover mapping applications. Pixel-based embeddings are typically distributed in raster file formats like GeoTIFF, but can also be found in NumPy or PyTorch file formats, lacking geospatial metadata and limiting use outside of the Python ecosystem. We struggled with the Google Satellite Embeddings due to upside down rasters, requiring reprojection before standard tools like rasterio or QGIS could load them.

\textbf{Reproducibility:} While most data and models are open source, many GitHub repositories are unmaintained and lack critical tools for reproducibility. The majority of model architectures are provided as is, with no unit tests or continuous integration to ensure that they work and no software tooling to make installation and usage easier. Many data sources were scraped from Microsoft's Planetary Computer or Google Earth Engine, and while the download locations are publicly available, these resources are continuously updated with new and improved versions of data products, meaning that the exact images used during model training may no longer be available.

\textbf{Generating New Embeddings:} Existing embeddings like Google Satellite Embedding have excellent spatiotemporal coverage. However, many embeddings only have partial coverage, and users may want to convert existing datasets like EuroSAT~\cite{eurosat} to embedded versions. With all foundation models distributed in different GitHub and Hugging Face repositories, this can be a time-intensive process. Other than the Clay and Presto embeddings, few products offer documented tutorials on generating new embeddings with the same models.


\section{Standardizing Access}\label{sec:torchgeo}

To resolve the difficulties of data loading, model access, and reproducibility, we added several new features to the TorchGeo library~\cite{torchgeo}. Previously, reproducing these experiments required stitching together four or more repositories and writing custom loaders. With TorchGeo, this becomes 20 lines of code.

TorchGeo already has model code and weights for ResNet-50 (DINO~\cite{dino}, MoCo~\cite{moco}, DeCUR~\cite{decur}), DINOv2 (SoftCon~\cite{softcon}), and Copernicus-FM~\cite{copernicus}. Likewise, TorchGeo already has data loaders for SSL4EO-S12~\cite{ssl4eo}, Copernicus-Pretrain~\cite{copernicus}, MMEarth~\cite{mmearth}, NLCD~\cite{nlcd}, GBIF, NAIP, Sentinel~1/2, and Landsat~8/9. During this project, we added support for the Tessera~\cite{tessera} and Presto~\cite{presto} models and data loaders for all known embedding products. This includes data loaders designed for search/retrieval (Clay, Major TOM, Earth Index) and for dense prediction tasks like land cover mapping (Copernicus, Presto, Tessera, Google).

Listing~\ref{lst:search} demonstrates an example of a search/retrieval task using TorchGeo. Cosine similarity is used as a distance metric, and we load the same model architecture and weights that were used to generate the Earth Index Embeddings. We grab a sample Sentinel-2 image and compute its embedding. Then we iterate over the Earth Index Embeddings and search for one (or more) locations with similar embeddings. These locations can then feed back into the original Sentinel-2 dataset, allowing for quick visualization of similar locations.

\begin{listing}[htbp]
\begin{minted}{python}
from torch.nn import CosineSimilarity
from torchgeo.datasets import (
    EarthIndexEmbeddings, Sentinel2
)
from torchgeo.models import (
    ViTSmall14_DINOv2_Weights, vit_small_patch14_dinov2,
)

cos = CosineSimilarity(dim=0)
model = vit_small_patch14_dinov2(
    ViTSmall14_DINOv2_Weights.SENTINEL2_ALL_SOFTCON
)
s2 = Sentinel2(paths)
sample = s2[xmin:xmax, ymin:ymax]
embed = model(sample)
eie = EarthIndexEmbeddings(root)
similarity = -1
x, y = None
for sample in eie:
    new_similarity = cos(embed, sample['embedding'])
    if new_similarity > similarity:
        similarity = new_similarity
        x, y = sample['x'], sample['y']

radius = 128 * 10  # pixels to meters
sample = s2[x-radius:x+radius, y-radius:y+radius]
s2.plot(sample)
\end{minted}
\caption{Example of search/retrieval task in TorchGeo.}
\label{lst:search}
\end{listing}

Listing~\ref{lst:map} shows an example of land cover mapping using the Google Satellite Embedding and EuroCrops~\cite{eurocrops} datasets. The Google Satellite Embedding dataset takes a list of one or more files or directories, and can read from local disk or remote cloud storage. The EuroCrops dataset automatically rasterizes vector shapefiles of crop type maps across Europe, and can be automatically downloaded. We compute the spatiotemporal intersection of these datasets to ensure that we have embeddings and masks from the same location and year. We then use gridded sampling of \(256 \times 256\) patches to ensure coverage of the entire continent. \(k\)-NN can be used to make predictions on unseen regions using knowledge extracted from embeddings and similarity with known agricultural fields.

\begin{listing}[htbp]
\begin{minted}{python}
from torch.utils.data import DataLoader
from torchgeo.datasets import (
    GoogleSatelliteEmbedding, EuroCrops
)
from torchgeo.samplers import GridGeoSampler

gse = GoogleSatelliteEmbedding(paths)
ec = EuroCrops(paths, download=True)
dataset = gse & ec  # spatiotemporal intersection
sampler = GridGeoSampler(dataset, size=256)
dataloader = DataLoader(dataset, sampler=sampler)

for batch in dataloader:
    # perform k-NN clustering using sklearn
\end{minted}
\caption{Example of land cover mapping task in TorchGeo.}
\label{lst:map}
\end{listing}

By standardizing the data loading process and treating embeddings as first-class datasets, we can ease support for users already familiar with TorchGeo-based workflows. 
TorchGeo also allows us to enable fair, side-by-side benchmarking of different embedding models on the same downstream tasks, forming the basis for future experiments.

\section{Discussion and Conclusion}\label{sec:conclusion}

In this work, we address the fragmentation of the Earth embedding ecosystem. By providing a comprehensive taxonomy (Section~\ref{sec:landscape}), landscape review (Section~\ref{sec:ecosystem}), and integration with TorchGeo (Section~\ref{sec:torchgeo}), we aim to ease the transition between product release and downstream application.

\textbf{Future Embedding Design Principles}: Based on our review, we strongly encourage future embedding releases to adopt the following design choices, critical for improving usability, comparability, and long-term impact.

\begin{tcolorbox}[
  colback=gray!10,
  colframe=black!40,
  boxrule=0.4pt,
  arc=2pt,
  left=5pt,
  right=5pt,
  top=5pt,
  bottom=5pt
]
\footnotesize
  \setlength{\itemsep}{0.3em}
  \textbf{Increase Input Data Diversity}: Sentinel~1/2 are useful but overrepresented. Consider other underexplored modalities such as the oceans and atmosphere. Hyperspectral data would be particularly promising, often requiring dimensionality reduction before usage anyway.

  \vspace{1ex}
  \textbf{Improve Explainability}: There is substantial room for improvements in uncertainty quantification and explainability in embedding products, including better provenance tracking. While raw imagery offers explainability with physical measurements and cloud masks, embeddings lose this information, making xAI challenging.

  \vspace{1ex}
  \textbf{Adopt Cloud-Native Formats}: Avoid distributing embeddings as raw \texttt{.pt} or \texttt{.npy} files. Prefer cloud-optimized formats (e.g., COG, GeoZarr, GeoParquet) that preserve geospatial metadata (CRS, transform).

  \vspace{1ex}
  \textbf{Standardize Benchmarks}: Many influential works either release new benchmarks or do not benchmark at all, making it difficult for practitioners to compare embedding performance. Using existing standardized benchmarks and releasing model weights enables fair comparison.
\end{tcolorbox}

Future work will focus on benchmark comparison, especially for time-series applications, and integrating more emerging models and embeddings into our unified interface.

\newpage
\small
\bibliographystyle{IEEEtranN}
\bibliography{references}

\begin{thebibliography}{33}
\providecommand{\natexlab}[1]{#1}
\providecommand{\url}[1]{#1}
\csname url@samestyle\endcsname
\providecommand{\newblock}{\relax}
\providecommand{\bibinfo}[2]{#2}
\providecommand{\BIBentrySTDinterwordspacing}{\spaceskip=0pt\relax}
\providecommand{\BIBentryALTinterwordstretchfactor}{4}
\providecommand{\BIBentryALTinterwordspacing}{\spaceskip=\fontdimen2\font plus
\BIBentryALTinterwordstretchfactor\fontdimen3\font minus \fontdimen4\font\relax}
\providecommand{\BIBforeignlanguage}[2]{{%
\expandafter\ifx\csname l@#1\endcsname\relax
\typeout{** WARNING: IEEEtranN.bst: No hyphenation pattern has been}%
\typeout{** loaded for the language `#1'. Using the pattern for}%
\typeout{** the default language instead.}%
\else
\language=\csname l@#1\endcsname
\fi
#2}}
\providecommand{\BIBdecl}{\relax}
\BIBdecl

\bibitem[Zhu et~al.(2026)Zhu, Xiong, Wang, Stewart, Heidler, Wang, Yuan, Dujardin, Xu, and Shi]{zhu2024foundations}
X.~X. Zhu, Z.~Xiong, Y.~Wang, A.~J. Stewart, K.~Heidler, Y.~Wang, Z.~Yuan, T.~Dujardin, Q.~Xu, and Y.~Shi, ``On the foundations of {Earth} foundation models,'' \emph{Nature Communications Earth \& Environment}, 2026.

\bibitem[Brown et~al.(2025)Brown, Kazmierski, Pasquarella, Rucklidge, Samsikova, Zhang, Shelhamer, Lahera, Wiles, Ilyushchenko, et~al.]{alphaearth}
C.~F. Brown, M.~R. Kazmierski, V.~J. Pasquarella, W.~J. Rucklidge, M.~Samsikova, C.~Zhang, E.~Shelhamer, E.~Lahera, O.~Wiles, S.~Ilyushchenko \emph{et~al.}, ``{AlphaEarth Foundations}: An embedding field model for accurate and efficient global mapping from sparse label data,'' \emph{arXiv preprint arXiv:2507.22291}, 2025.

\bibitem[Clay(2024)]{clay}
Clay, ``Clay foundation model,'' https://huggingface.co/made-with-clay/Clay, 2024.

\bibitem[Gomes et~al.(2025)Gomes, Wittmann, Robert, Jakubik, Reichelt, Maurogiovanni, Vinge, Hurst, Scheurer, Sedona, et~al.]{gomes2025lossy}
C.~Gomes, I.~Wittmann, D.~Robert, J.~Jakubik, T.~Reichelt, S.~Maurogiovanni, R.~Vinge, J.~Hurst, E.~Scheurer, R.~Sedona \emph{et~al.}, ``Lossy neural compression for geospatial analytics: A review,'' \emph{IEEE Geoscience and Remote Sensing Magazine}, 2025.

\bibitem[Rolf et~al.(2025)Rolf, Klemmer, and Ru{\ss}wurm]{rolf2025earth}
E.~Rolf, K.~Klemmer, and M.~Ru{\ss}wurm, ``Earth embeddings: Harnessing the information in {Earth} observation data with machine learning,'' in \emph{Proceedings of the Special Interest Group on Computer Graphics and Interactive Techniques Conference Frontiers}, 2025, pp. 1--2.

\bibitem[Klemmer et~al.(2025{\natexlab{a}})Klemmer, Rolf, Russwurm, Camps-Valls, Czerkawski, Ermon, Francis, Jacobs, Kerner, Mackey, et~al.]{klemmer2025earth}
K.~Klemmer, E.~Rolf, M.~Russwurm, G.~Camps-Valls, M.~Czerkawski, S.~Ermon, A.~Francis, N.~Jacobs, H.~R. Kerner, L.~Mackey \emph{et~al.}, ``Earth embeddings: Towards {AI}-centric representations of our planet,'' 2025, earthArXiv.

\bibitem[Stewart et~al.(2025)Stewart, Robinson, Corley, Ortiz, Lavista~Ferres, and Banerjee]{torchgeo}
A.~J. Stewart, C.~Robinson, I.~A. Corley, A.~Ortiz, J.~M. Lavista~Ferres, and A.~Banerjee, ``{TorchGeo}: Deep learning with geospatial data,'' \emph{ACM Transactions on Spatial Algorithms and Systems}, vol.~11, no.~4, pp. 1--28, Aug. 2025.

\bibitem[Francis and Czerkawski(2024)]{majortom}
A.~Francis and M.~Czerkawski, ``{Major TOM}: Expandable datasets for {Earth} observation,'' in \emph{IGARSS 2024-2024 IEEE International Geoscience and Remote Sensing Symposium}.\hskip 1em plus 0.5em minus 0.4em\relax IEEE, 2024, pp. 2935--2940.

\bibitem[{Earth Genome}(2025)]{earthindex}
{Earth Genome}, ``Earth index embeddings,'' https://source.coop/earthgenome/earthindexembeddings, 2025.

\bibitem[Wang et~al.(2025)Wang, Xiong, Liu, Stewart, Dujardin, Bountos, Zavras, Gerken, Papoutsis, Leal-Taix{\'e}, and Zhu]{copernicus}
Y.~Wang, Z.~Xiong, C.~Liu, A.~J. Stewart, T.~Dujardin, N.~I. Bountos, A.~Zavras, F.~Gerken, I.~Papoutsis, L.~Leal-Taix{\'e}, and X.~X. Zhu, ``Towards a unified {Copernicus} foundation model for {Earth} vision,'' in \emph{Proceedings of the IEEE/CVF International Conference on Computer Vision}, Oct. 2025, pp. 9888--9899.

\bibitem[Zvonkov et~al.(2025)Zvonkov, Tseng, Becker-Reshef, and Kerner]{prestoembed}
I.~Zvonkov, G.~Tseng, I.~Becker-Reshef, and H.~Kerner, ``Cropland mapping using geospatial embeddings,'' \emph{arXiv preprint arXiv:2511.02923}, 2025.

\bibitem[Feng et~al.(2025)Feng, Atzberger, Jaffer, Knezevic, Sormunen, Young, Lisaius, Immitzer, Jackson, Ball, et~al.]{tessera}
Z.~Feng, C.~Atzberger, S.~Jaffer, J.~Knezevic, S.~Sormunen, R.~Young, M.~C. Lisaius, M.~Immitzer, T.~Jackson, J.~Ball \emph{et~al.}, ``{TESSERA}: Temporal embeddings of surface spectra for {Earth} representation and analysis,'' \emph{arXiv preprint arXiv:2506.20380}, 2025.

\bibitem[Xiong et~al.(2024)Xiong, Wang, Zhang, Stewart, Hanna, Borth, Papoutsis, Le~Saux, Camps-Valls, and Zhu]{dofa}
Z.~Xiong, Y.~Wang, F.~Zhang, A.~J. Stewart, J.~Hanna, D.~Borth, I.~Papoutsis, B.~Le~Saux, G.~Camps-Valls, and X.~X. Zhu, ``Neural plasticity-inspired multimodal foundation model for {Earth} observation,'' Mar. 2024, arXiv preprint arXiv:2403.15356.

\bibitem[Herzog et~al.(2025)Herzog, Bastani, Zhang, Tseng, Redmon, Sablon, Park, Morrison, Buraczynski, Farley, et~al.]{olmoearth}
H.~Herzog, F.~Bastani, Y.~Zhang, G.~Tseng, J.~Redmon, H.~Sablon, R.~Park, J.~Morrison, A.~Buraczynski, K.~Farley \emph{et~al.}, ``{OlmoEarth}: Stable latent image modeling for multimodal {Earth} observation,'' \emph{arXiv preprint arXiv:2511.13655}, 2025.

\bibitem[Klemmer et~al.(2025{\natexlab{b}})Klemmer, Rolf, Robinson, Mackey, and Ru{\ss}wurm]{klemmer2025satclip}
K.~Klemmer, E.~Rolf, C.~Robinson, L.~Mackey, and M.~Ru{\ss}wurm, ``{SatCLIP}: Global, general-purpose location embeddings with satellite imagery,'' in \emph{Proceedings of the AAAI Conference on Artificial Intelligence}, vol.~39, no.~4, 2025, pp. 4347--4355.

\bibitem[Ru{\ss}wurm et~al.(2023)Ru{\ss}wurm, Klemmer, Rolf, Zbinden, and Tuia]{russwurm2023geographic}
M.~Ru{\ss}wurm, K.~Klemmer, E.~Rolf, R.~Zbinden, and D.~Tuia, ``Geographic location encoding with spherical harmonics and sinusoidal representation networks,'' \emph{arXiv preprint arXiv:2310.06743}, 2023.

\bibitem[Vinge et~al.(2025)Vinge, Wittmann, Schneider, Marszalek, Gilch, Brunschwiler, and Albrecht]{vinge2025neuco}
R.~Vinge, I.~Wittmann, J.~Schneider, M.~Marszalek, L.~Gilch, T.~Brunschwiler, and C.~M. Albrecht, ``{NeuCo-Bench}: A novel benchmark framework for neural embeddings in {Earth} observation,'' \emph{arXiv preprint arXiv:2510.17914}, 2025.

\bibitem[Rao et~al.(2025)Rao, Ru{\ss}wurm, Klemmer, and Rolf]{rao2025measuring}
A.~Rao, M.~Ru{\ss}wurm, K.~Klemmer, and E.~Rolf, ``Measuring the intrinsic dimension of {Earth} representations,'' \emph{arXiv preprint arXiv:2511.02101}, 2025.

\bibitem[Albrecht et~al.(2025)Albrecht, Schneider, Vinge, and Wittmann]{albrecht20252025}
C.~Albrecht, J.~Schneider, R.~Vinge, and I.~Wittmann, ``The 2025 {CVPR EARTHVISION} data challenge by {Embed2Scale},'' in \emph{IEEE/CVF Conference on Computer Vision and Pattern Recognition}, 2025.

\bibitem[Jakubik et~al.(2025)Jakubik, Yang, Blumenstiel, Scheurer, Sedona, Maurogiovanni, Bosmans, Dionelis, Marsocci, Kopp, et~al.]{jakubik2025terramind}
J.~Jakubik, F.~Yang, B.~Blumenstiel, E.~Scheurer, R.~Sedona, S.~Maurogiovanni, J.~Bosmans, N.~Dionelis, V.~Marsocci, N.~Kopp \emph{et~al.}, ``{TerraMind}: Large-scale generative multimodality for {Earth} observation,'' \emph{arXiv preprint arXiv:2504.11171}, 2025.

\bibitem[Pettersson and Daoud(2025)]{pettersson2025leveraging}
M.~B. Pettersson and A.~Daoud, ``Leveraging compact satellite embeddings and graph neural networks for large-scale poverty mapping,'' \emph{arXiv preprint arXiv:2511.01408}, 2025.

\bibitem[He et~al.(2016)He, Zhang, Ren, and Sun]{resnet}
K.~He, X.~Zhang, S.~Ren, and J.~Sun, ``Deep residual learning for image recognition,'' in \emph{Proceedings of the IEEE Conference on Computer Vision and Pattern Recognition}, 2016, pp. 770--778.

\bibitem[Woo et~al.(2023)Woo, Debnath, Hu, Chen, Liu, Kweon, and Xie]{convnext}
S.~Woo, S.~Debnath, R.~Hu, X.~Chen, Z.~Liu, I.~S. Kweon, and S.~Xie, ``{ConvNeXt V2}: Co-designing and scaling {ConvNets} with masked autoencoders,'' in \emph{Proceedings of the IEEE/CVF Conference on Computer Vision and Pattern Recognition}, 2023, pp. 16\,133--16\,142.

\bibitem[Helber et~al.(2019)Helber, Bischke, Dengel, and Borth]{eurosat}
P.~Helber, B.~Bischke, A.~Dengel, and D.~Borth, ``{EuroSAT}: A novel dataset and deep learning benchmark for land use and land cover classification,'' \emph{IEEE Journal of Selected Topics in Applied Earth Observations and Remote Sensing}, vol.~12, no.~7, pp. 2217--2226, 2019.

\bibitem[Caron et~al.(2021)Caron, Touvron, Misra, J{\'e}gou, Mairal, Bojanowski, and Joulin]{dino}
M.~Caron, H.~Touvron, I.~Misra, H.~J{\'e}gou, J.~Mairal, P.~Bojanowski, and A.~Joulin, ``Emerging properties in self-supervised vision transformers,'' in \emph{Proceedings of the IEEE/CVF International Conference on Computer Vision}, 2021, pp. 9650--9660.

\bibitem[Chen et~al.(2020)Chen, Fan, Girshick, and He]{moco}
X.~Chen, H.~Fan, R.~Girshick, and K.~He, ``Improved baselines with momentum contrastive learning,'' \emph{arXiv preprint arXiv:2003.04297}, 2020.

\bibitem[Wang et~al.(2024{\natexlab{a}})Wang, Albrecht, Ait Ali~Braham, Liu, Xiong, and Zhu]{decur}
Y.~Wang, C.~M. Albrecht, N.~Ait Ali~Braham, C.~Liu, Z.~Xiong, and X.~X. Zhu, ``Decoupling common and unique representations for multimodal self-supervised learning,'' in \emph{European Conference on Computer Vision}.\hskip 1em plus 0.5em minus 0.4em\relax Springer, 2024, pp. 286--303.

\bibitem[Wang et~al.(2024{\natexlab{b}})Wang, Albrecht, and Zhu]{softcon}
Y.~Wang, C.~M. Albrecht, and X.~X. Zhu, ``Multi-label guided soft contrastive learning for efficient {Earth} observation pretraining,'' \emph{IEEE Transactions on Geoscience and Remote Sensing}, 2024.

\bibitem[Wang et~al.(2023)Wang, Ait Ali~Braham, Xiong, Liu, Albrecht, and Zhu]{ssl4eo}
Y.~Wang, N.~Ait Ali~Braham, Z.~Xiong, C.~Liu, C.~M. Albrecht, and X.~X. Zhu, ``{SSL4EO-S12}: A large-scale multimodal, multitemporal dataset for self-supervised learning in {Earth} observation,'' \emph{IEEE Geoscience and Remote Sensing Magazine}, vol.~11, no.~3, pp. 98--106, 2023.

\bibitem[Nedungadi et~al.(2024)Nedungadi, Kariryaa, Oehmcke, Belongie, Igel, and Lang]{mmearth}
V.~Nedungadi, A.~Kariryaa, S.~Oehmcke, S.~Belongie, C.~Igel, and N.~Lang, ``{MMEarth}: Exploring multi-modal pretext tasks for geospatial representation learning,'' in \emph{European Conference on Computer Vision}.\hskip 1em plus 0.5em minus 0.4em\relax Springer, 2024, pp. 164--182.

\bibitem[Sohl et~al.(2025)Sohl, Jin, Dewitz, Wickham, Brown, Stehman, Herold, Schleeweis, Tollerud, and Deering]{nlcd}
T.~Sohl, S.~Jin, J.~Dewitz, J.~Wickham, J.~Brown, S.~Stehman, N.~Herold, K.~Schleeweis, H.~Tollerud, and C.~Deering, ``Thirty years of the {US National Land Cover Database}: impacts and future direction,'' \emph{Photogrammetric Engineering \& Remote Sensing}, vol.~91, no.~10, pp. 647--659, 2025.

\bibitem[Tseng et~al.(2023)Tseng, Cartuyvels, Zvonkov, Purohit, Rolnick, and Kerner]{presto}
G.~Tseng, R.~Cartuyvels, I.~Zvonkov, M.~Purohit, D.~Rolnick, and H.~Kerner, ``Lightweight, pre-trained transformers for remote sensing timeseries,'' \emph{arXiv preprint arXiv:2304.14065}, 2023.

\bibitem[Schneider et~al.(2023)Schneider, Schelte, Schmitz, and K{\"o}rner]{eurocrops}
M.~Schneider, T.~Schelte, F.~Schmitz, and M.~K{\"o}rner, ``{EuroCrops}: The largest harmonized open crop dataset across the {European Union},'' \emph{Scientific Data}, vol.~10, no.~1, p. 612, 2023.

\end{thebibliography}

\end{document}